\documentclass[floatfix,showpacs,amsmath,amsfonts,amssymb,aps,twocolumn,
superscriptaddress,prl,10pt]{revtex4-1}
\usepackage[pdftex]{graphicx}

\usepackage{color}

\newcommand{\figref}[1]{Fig.~\ref{#1}}
\newcommand{\e}[1]{\text{e}^{#1}}
\newcommand{\cmplxi}{\text{i}}
\newcommand{\diffd}{\text{d}}
\renewcommand{\vec}[1]{\mathbf{#1}}
\newcommand{\punc}[1]{\,#1}
\newcommand{\neweqnline}{\nonumber\\}

\newcommand{\eqnref}[1]{Eqn.~(\ref{#1})}

\begin{document}

\title{Pseudopotential for the electron-electron interaction}
\author{J.H.~Lloyd-Williams}
\affiliation{Cavendish~Laboratory, J.J.~Thomson~Avenue, Cambridge, CB3~0HE, 
  United Kingdom}
\author{R.J.~Needs}
\affiliation{Cavendish~Laboratory, J.J.~Thomson~Avenue, Cambridge, CB3~0HE, 
  United Kingdom}
\author{G.J.~Conduit}
\affiliation{Cavendish~Laboratory, J.J.~Thomson~Avenue, Cambridge, CB3~0HE, 
  United Kingdom}
\date{\today}

\begin{abstract}
  We propose a pseudopotential for the electron-electron Coulomb interaction
  to improve the efficiency of many-body electronic structure
  calculations. The pseudopotential accurately replicates the scattering
  properties of the Coulomb interaction, and recovers the analytical
  solution for two electrons in a parabolic trap. A case study for the
  homogeneous electron gas using the diffusion Monte Carlo and configuration
  interaction methods recovers highly accurate values for the ground state
  energy, and the smoother potential reduces the computational cost by a
  factor of $\sim30$.  Finally, we demonstrate the use of the
  pseudopotential to study isolated lithium and beryllium atoms.
\end{abstract}


\maketitle

Electron-electron interactions drive chemical reactions, govern material
properties, and conspire to form strongly correlated phases. Despite the
widespread and important consequences of electronic correlations, leading
computational techniques such as diffusion Monte Carlo
(DMC)~\cite{Foulkes_2001}, truncated configuration interaction
(CI)~\cite{Hylleraas28,Meyer76}, M{\o}ller-Plesset theory~\cite{Moller34},
coupled cluster theory~\cite{Cizek66}, and F12 methods~\cite{Valeev12}.
These approaches are very expensive for real-life systems because the
divergence in the electron-electron Coulomb interaction must be sampled
finely~\cite{Hill85,Kutzelnigg92}. Here we propose a pseudopotential that
accurately replicates the scattering properties of the Coulomb interaction,
delivers the ground state energies within chemical accuracy of
$1$\,kcal\,mol$^{-1}$, but does not diverge, which reduces the computational cost
of both DMC and CI by a factor of $\sim 30$.

Pseudopotentials were first introduced by Hellmann~\cite{Hellmann35} to
describe the attractive electron-ion interaction in molecules and
solids. Integrating out the core electrons that screen the central ion
leaves a pseudopotential for the valence electrons. The reduction in the
number of electrons and the greater smoothness of the electron-ion
pseudopotential provides computational advantages that led to their
widespread adoption in electronic structure calculations, including density
functional theory~\cite{Bachelet82} and DMC methods~\cite{Trail05}.

First principles approaches must still account for the divergent repulsive
electron-electron interaction that necessitates fine
sampling~\cite{Hill85,Kutzelnigg92}. The Kato cusp conditions~\cite{Kato57,
  Pack66,Boys60,Drummond04,Kutzelnigg85,Klopper87,Kutzelnigg91,Gruneis12}
enforce a wavefunction with a kinetic energy divergence that cancels the
Coulomb divergence, leaving a remnant finite discontinuity in the local
energy $\psi^{-1}(\vec{R})\hat{H}\psi(\vec{R})$, which is evaluated with the
electrons at point $\vec{R}$ in configuration space. There have been
attempts to apply a local density solution to the short-ranged
behavior~\cite{Henderson04,Henderson05}. It was also proposed to introduce a
soft-Coulomb operator either in real space~\cite{Clementi65}, or reciprocal
space~\cite{Panas95}.  Another attempt was to split the Coulomb interaction
into a short and long-ranged component, so that they could be handled
separately~\cite{Savin96}.  However, at present pseudopotentials are not
generally used to smooth the electron-electron interaction.

We develop an accurate electron-electron pseudopotential for electrons
scattering with any energy and angular momentum. We build on the formalism
used to construct a pseudopotential for the contact interaction found in
ultracold atomic gases~\cite{Bugnion14}. This formalism is somewhat
different from the standard pseudopotential approach developed for
attractive electron-ion interactions that focuses on discrete bound state
energies~\cite{Gilbert63,King96,Nooijen98,Prendergast01}, although it can be
extended to scattering states~\cite{Hamann89}. The proposed pseudopotential
is identical to the Coulomb interaction outside of a cut-off radius where
many-body physics becomes important. The pseudopotential delivers all of the
physics of the Coulomb interaction but does not diverge, so that the ground state
can be determined efficiently. After developing the pseudopotential in the
two-body scattering problem, we test it on the analytically solvable system
of two electrons in a parabolic trap~\cite{Taut93}.

We study the applicability, accuracy, and portability of the pseudopotential
for a homogeneous electron gas (HEG) using two methods: DMC in which the use
of the pseudopotential reduces the required time-step, and CI in which the
pseudopotential reduces the size of the plane-wave basis set required. The
pseudopotential delivers chemical accuracy, and at the same time reduces the
computational cost of both techniques by a factor of $\sim30$. Finally, we
test the pseudopotential on two inhomogeneous systems, the isolated lithium
and beryllium atoms.

\section{Construction of the pseudopotential}

\begin{figure}
 \begin{tabular}{ll}
  (a) Pseudopotential&(b) Error with cutoff\\
  \includegraphics[width=0.48\linewidth]{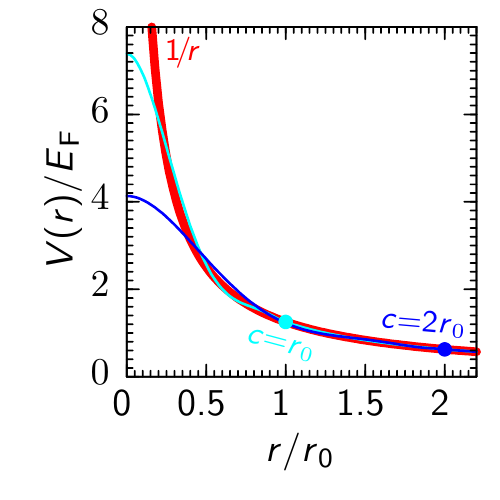}&
  \includegraphics[width=0.48\linewidth]{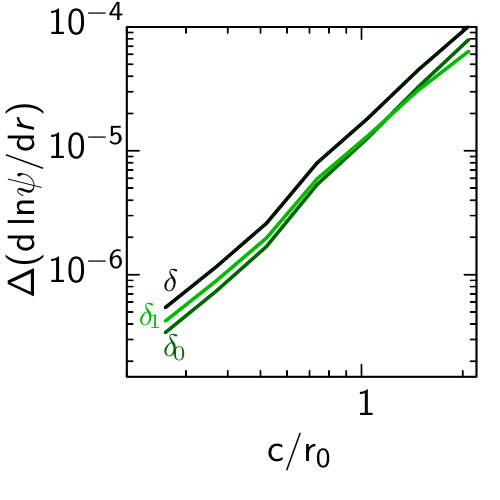}\\
  (c) Error with density&(d) Two-body scattering\\
  \includegraphics[width=0.48\linewidth]{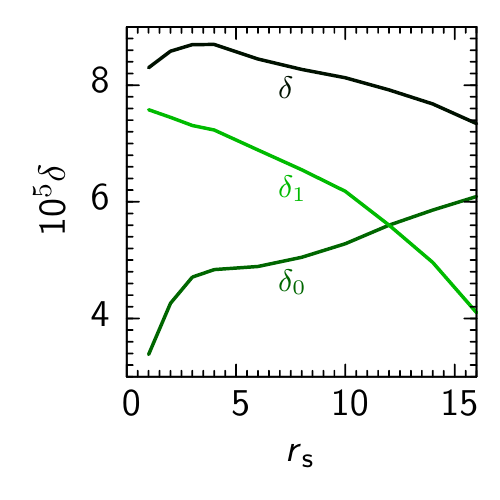}&
  \includegraphics[width=0.48\linewidth]{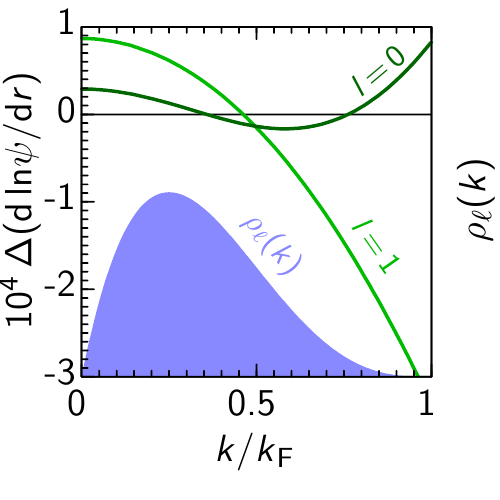}
 \end{tabular}
 \caption{(Color online)
   (a) The interaction potentials: the Coulomb potential is shown in red,
   the pseudopotential with cutoff radius $c=r_{0}$ is in cyan, and
   the pseudopotential with cutoff radius $c=2r_{0}$ is in blue.
   (b) The error in the logarithmic derivative of the scattering wavefunction
   with cutoff radius for an electron gas with $r_{\text{s}}=2$.
   $\delta$ shows the error summed over angular momentum channels,
   $\delta_{0}$ is the contribution from the $\ell=0$ channel and
   $\delta_{1}$ from the $\ell=1$ channel.
   (c) The pseudopotential error for a range of $r_{\text{s}}$ values.
   (d) The error in the logarithmic derivative of the scattering
   wavefunction with incident wave vector for the $\ell=0$ and $\ell=1$
   scattering channels. The filled blue curve plotted on an arbitrary linear
   scale on the secondary y-axis shows the weighting factor $\rho_{\ell}(k)$
   used in evaluating the overall error in the logarithmic derivative.
 }
 \label{fig:Pseudopotential}
\end{figure}

To construct the pseudopotential, we adopt the formalism of
Ref.~\cite{Bugnion14} and study the two-body problem: two electrons in their
center-of-mass frame with wave vector $k\ge0$ and angular momentum quantum
number $\ell$. The Hamiltonian in atomic units is
$-\frac{1}{r^2}\frac{\diffd}{\diffd r}(r^2\frac{\diffd\psi}{\diffd
  r})+\frac{\ell(\ell+1)}{r^2}\psi+V(r)\psi=k^2\psi$, and the repulsive
Coulomb potential is $V(r)=1/r$.  The proposed pseudopotential is identical
to the Coulomb potential outside of a cutoff radius $c$, and at the cutoff
it is continuous and has a continuous first derivative. At small
electron-electron separation $r$, the pseudopotential can be chosen to be
softer than the Coulomb interaction so that on electron coalescence at $r=0$
it is finite and has zero gradient to remove possible divergences and
discontinuities in the local energy, thereby reducing the variance in our
estimate of the total energy. These considerations suggest a pseudopotential of
the form
\begin{align}
 V\!(r)\!=\!\frac{1}{c}\!
 \begin{cases}
  \begin{aligned}
   &1+\left(1-\frac{r}{c}\right)\left(\frac{r}{c}\right)^{\!2}+\\
   &\!\!\left(1\!-\!\frac{r}{c}\right)^{\!2}\!\!
   \left[v_{1}\!\left(\!\frac{1}{2}\!+\!\frac{r}{c}\!\right)\!+\!\!
   \sum_{i=2}^{N_{\text{v}}}v_{i}\!\left(\frac{r}{c}\right)^{\!i}\!\right]
  \end{aligned}&0\le r\le c\\
  \begin{aligned}
   \frac{c}{r}
  \end{aligned}&r>c\punc{,}
 \end{cases}
 \nonumber
\end{align}
with variational freedom introduced by a polynomial expansion of order
$N_{\text{v}}=6$. To determine the parameters $\{v_{i}\}$ we calculate the
scattering states. The scattering states $\psi_{k,\ell}(r)$ for the Coulomb
interaction can be solved exactly in terms of Whittaker functions, whereas
the scattering solution $\phi_{k,\ell}(r)$ from the pseudopotential is
solved numerically. The difference between the scattering properties of the
two potentials is characterized by the mean square error in the logarithmic
derivative of the scattering wavefunction at the cutoff radius
\begin{align}
 \delta^{2}=\sum_{\ell=0}^{6}
  \int_{0}^{k_{\text{F}}}\!\!\!\!\!\diffd\vec{k}\rho_{\ell}(\vec{k})\!
 \left(\!
 \left.\frac{\diffd\ln\psi_{k,\ell}}{\diffd(r/c)}\right|_{r=c}\!\!\!-
 \left.\frac{\diffd\ln\phi_{k,\ell}}{\diffd(r/c)}\right|_{r=c}
 \right)^{\!2}\!\!\punc{,}
 \label{eqn:phaseshift}
\end{align}
which is summed over all angular momentum channels $\ell=\{0,\dots,6\}$ and
integrated over all possible scattering wave vectors $0\le k\le
k_{\text{F}}$ encountered in an electron gas with Fermi momentum
$k_{\text{F}}$~\cite{Prendergast01}. Following Ref.~\cite{Bugnion14}, we
weight the importance of different scattering states by a factor
$\rho_{\ell}(k)$, which is chosen to replicate the density of scattering
states in a Hartree-Fock trial wavefunction for a homogeneous electron gas
where $\rho_{\ell}(\vec{k})=
\int\diffd\vec{q}\,n_{\text{F}}(\vec{q})n_{\text{F}}(\vec{k}+\vec{q})/
\sqrt{(2\ell+1)!!}$, and $n_{\text{F}}$ is the Fermi-Dirac distribution
function. We select the variational parameters $\{v_{i}\}$ that minimize
$\delta^2$, which gives a pseudopotential whose scattering closely
replicates the Coulomb interaction.  We associate the length scale
$r_{0}=(9\pi/4)^{1/3}/k_{\text{F}}$ with the typical electron separation and
characterize an electron gas with the standard density parameter
$r_{\text{s}}=r_{0}/a_{\text{B}}$, where $a_{\text{B}}$ is the electron Bohr
radius)

In \figref{fig:Pseudopotential}(a) we examine two of the pseudopotentials
constructed to be used in an electron gas with density $r_{\text{s}}=2$.  At
small $r$ the pseudopotential is flat to ensure that the wavefunction is
smooth.  The pseudopotential is therefore weaker than the Coulomb potential
but, to give the same net scattering strength, the pseudopotential must
exceed the Coulomb potential at intermediate $r$, before they merge at the
cutoff radius. The figure also shows that on reducing the cutoff radius the
pseudopotential approaches the Coulomb potential. Therefore, the
pseudopotential should recover the scattering properties of the Coulomb
potential with increasing accuracy as the pseudopotential cutoff radius is
reduced.  We verify this in \figref{fig:Pseudopotential}(b) where the error
in the logarithmic derivative of the scattering wavefunction falls with
cutoff radius as $\sim(c/r_{0})^{2.6}$. The $\ell=0$ and $\ell=1$ channels
provide similar contributions to the error in the scattering wave
function. Now that we have tested the pseudopotential developed for an
electron gas at $r_{\text{s}}=2$, we develop and test pseudopotentials to be
applied to electron gases with the full range of densities $1\le
r_{\text{s}}\le16$ that can be found in real-life systems.
\figref{fig:Pseudopotential}(c) shows that the average error in the
logarithmic derivative $\delta\lesssim10^{-4}$ is small compared with the
typical scattering phase shift $2\pi$ over a wide range of electron gas
densities, demonstrating that the pseudopotential accurately reproduces the
two-body scattering properties of the Coulomb interaction.

The error in the logarithmic derivative of the wavefunction averaged over
the incident wave vectors of electrons scattering off the pseudopotential is
small.  To understand how this is achieved, we examine in
\figref{fig:Pseudopotential}(d) the variation of the error in the
logarithmic derivative with respect to the incident wave vector. The
$\ell=0$ channel has a quadratic form that crosses zero error twice, whereas
the $\ell=1$ channel has an error that crosses zero only once. The
variational freedom in the pseudopotential has been used to minimize the
error around $k\approx0.3k_{\text{F}}$ where the density of scattering
states is largest, sacrificing accuracy at higher incident wave vectors.

With the pseudopotential providing phase shifts with an error of only
$\sim10^{-4}$, we are well-positioned to test its performance in a many-body
setting. We first study an idealized system with an analytical solution to
provide an exact benchmark: two electrons in a parabolic trap. We also study
systems that cannot be solved analytically: the HEG with two complementary
methods; DMC and CI; and we also study isolated lithium and beryllium atoms.
This allows us to assess the performance and accuracy of the
pseudopotential, and verify its portability.

\section{Two electrons in a parabolic trap}

\begin{figure}
 \begin{tabular}{ll}
  (a) $\ell\!=\!0$ wavefunction&(b) Two-body scattering\\
  \includegraphics[width=0.48\linewidth]{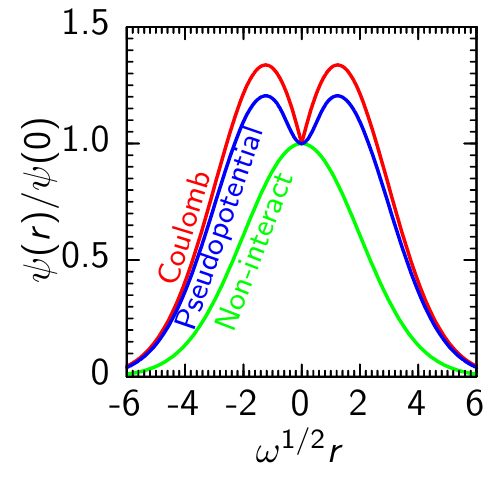}&
  \includegraphics[width=0.48\linewidth]{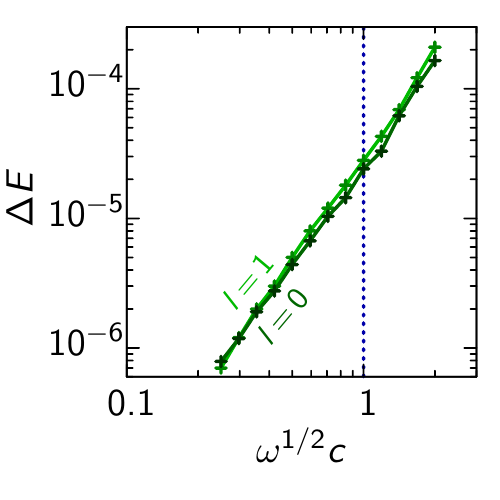}
 \end{tabular}
 \caption{(Color online)
   (a) Wavefunction of the relative motion of two electrons
   in a harmonic trap in the $\ell=0$ angular momentum state. 
   The green curve shows the wavefunction for non-interacting electrons, 
   the red shows electrons interacting via the Coulomb interaction, 
   and blue shows interactions via the pseudopotential.
   (b) The error per electron in the estimate of the ground state energy of two
   opposite-spin ($\ell=0$) and same-spin ($\ell=1$) electrons
   in a parabolic trap as a function of cutoff radius. The vertical
   blue dotted line shows the typical separation of the electrons in the harmonic trap.
 }
 \label{fig:Parabola}
\end{figure}

Now that we have constructed the Coulomb pseudopotential and calibrated it
against the phase shift of two atoms scattering in a vacuum, we evaluate the
accuracy of the pseudopotential in a second analytically soluble system:
Hooke's atom, two interacting electrons trapped in the parabolic well
$m\omega^2r^2/2$~\cite{Taut93}.  This problem received early numerical
attention~\cite{Kestner62,Laufer86,Merkt91}, and was more recently studied
with coupled cluster methods~\cite{Henderson01,Henderson03}.  We solve
separately for opposite- and same-spin electrons as the relative
wavefunctions differ due to fermion antisymmetry.

We solve for the energy of two interacting electrons in the parabolic trap
within the center-of-mass frame in which the interacting Hamiltonian for
relative motion is $\hat{H}=-\frac{1}{r^2}\frac{\diffd}{\diffd
  r}(r^2\frac{\diffd}{\diffd r})+\omega^2r^2/4+\ell(\ell+1)/r^2+V(r)$, where
$V(r)=1/r$ is the Coulomb interaction in atomic units. For the special case
of $\omega=1/8$ this model can be solved analytically for the $\ell=0$
(opposite-spin electrons) ground state giving eigenenergy $E=5/4$ (the
non-interacting center-of-mass Hamiltonian has energy $3/4$ giving a total
energy $E=2$). On replacing the interaction potential by a pseudopotential,
the Hamiltonian for relative motion can be solved numerically and the ground
state energy compared with the exact solution for the Coulomb
interaction. When constructing the pseudopotential we chose a maximum energy
of the scattering states that we integrate over in
\eqnref{eqn:phaseshift}. We take this to be the energy per electron in the
interacting system, $E=1$.

The parabolic trap is an ideal setting to compare the ground state
wavefunction predicted by the Coulomb interaction with that from the
pseudopotential.  In \figref{fig:Parabola}(a) we show the $\ell=0$ (i.e.,
opposite-spin electrons) ground state wavefunction for relative electron
motion.  Firstly, to orient the discussion we show the wavefunction for
non-interacting electrons, which is a Gaussian that is smooth at electron
coalescence. The wavefunction for the Coulomb interaction has a gradient
discontinuity at electron coalescence which provides a divergent kinetic
energy that cancels the divergence in the Coulomb interaction. In general
the gradient discontinuity is difficult to capture numerically and it
hinders computational approaches. However, the smooth pseudopotential
provides a wavefunction that is smooth over all space including at electron
coalescence, which should aid computational methods.

In \figref{fig:Parabola}(b) we study the error in the $\ell=0$ ground state
energy when varying the cutoff radius, which is the control parameter for
adjusting the accuracy of the pseudopotential. The error in the ground state
energy with the cutoff set to the typical electron separation,
$1/\sqrt{\omega}$, is $2\times10^{-5}$\,au per electron. With decreasing
cutoff radius $c$ the pseudopotential approaches the Coulomb interaction and
the accuracy further increases, varying as
$\sim(\sqrt{\omega}c)^{2.7}$. This scaling in error with cutoff radius is
similar to that seen in the error in the logarithmic derivative of the
scattering wavefunction shown in \figref{fig:Pseudopotential}(b), which
varies as $\sim(c/r_{0})^{2.6}$.

The interactions between opposite-spin and same-spin electrons both make
important contributions to the total energy in many systems. Therefore, we
next study the ground state energy of same-spin electrons in a parabolic
trap. This requires a spatially anti-symmetric ground state, and so we
require the system with $\ell=1$. Here the system is analytically soluble
with $\omega=1/16$ giving an energy of $E=21/16$.  In
\figref{fig:Parabola}(b) we study the error in the prediction of the
$\ell=1$ energy. The error in the ground state energy with the cutoff set to
the typical electron separation, $1/\sqrt{\omega}$, is $3\times10^{-5}$\,au
per electron. With decreasing cutoff radius $c$ the accuracy further
increases, varying as $\sim(\sqrt{\omega}c)^{2.55}$. The errors achieved for
both the $\ell=0$ and $\ell=1$ channels are two orders of magnitude better
than the target chemical accuracy of $1$\,kcal\,mol$^{-1}=0.0016\text{\,au}$
per electron.  The proposed pseudopotential is therefore sufficiently
accurate for scattering between both opposite- and same-spin electrons in
this two-body system.

\section{HEG with Diffusion Monte Carlo}

\begin{figure}
 \begin{tabular}{ll}
  (a) Time-step error&(b) Error with cutoff radius\\
  \includegraphics[width=0.45\linewidth]{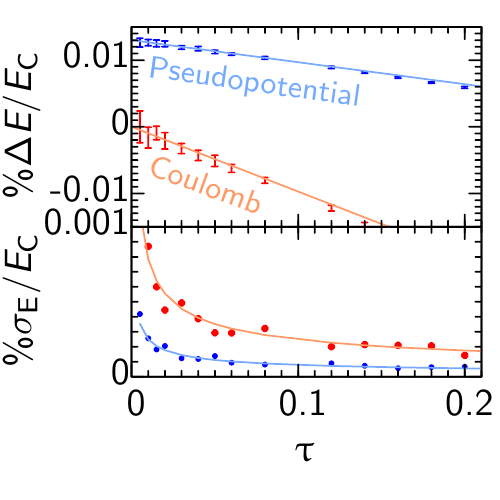}&
  \includegraphics[width=0.45\linewidth]{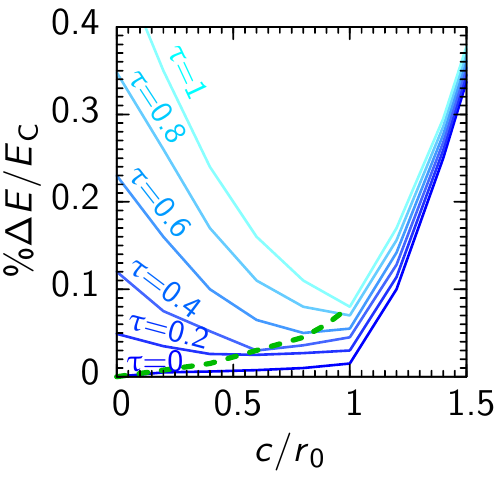}\\
  (c) Uncertainty with cutoff&(d) Pair correlation function\\
  \includegraphics[width=0.45\linewidth]{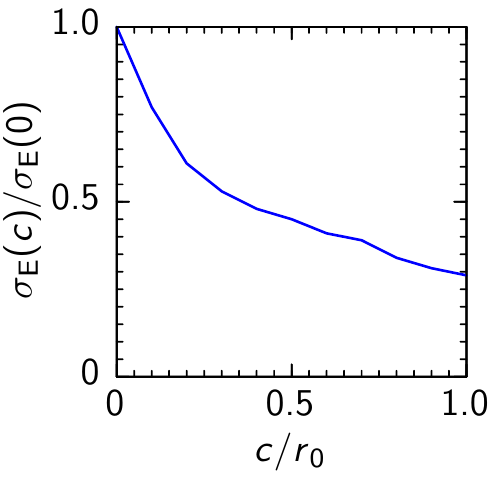}&
  \includegraphics[width=0.45\linewidth]{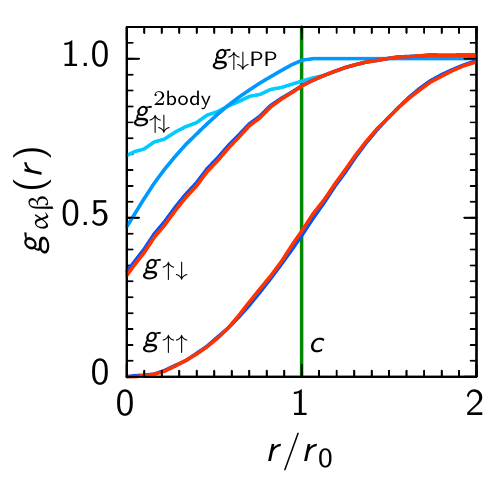}
 \end{tabular}
 (e) Error with density\hspace*{\fill}~\\
 \includegraphics[width=0.9\linewidth]{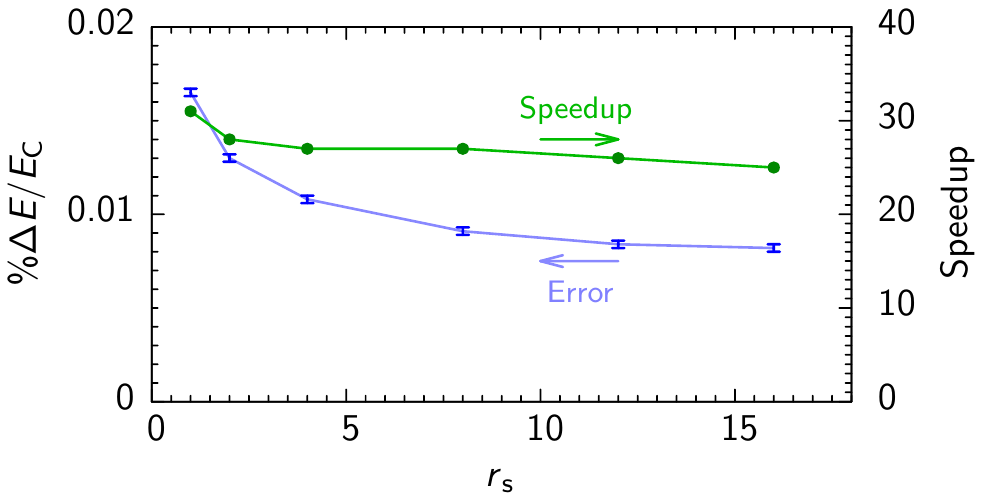}
 \caption{(Color online)
   (a)~Upper: The error in the energy of the HEG with DMC time step. 
   The y-axis origin is set by the extrapolation to zero time step 
   of the energy obtained with the Coulomb interaction. The red error bars 
   show the Coulomb interaction and the blue error bars the pseudopotential, 
   the solid lines show linear extrapolation to zero time step.
   Lower: The uncertainty in energy predictions, with the lines
   showing the $\tau^{-1/2}$ fit.
   (b)~The error in the energy with cutoff radius for different time steps 
   $\tau\in\{0,\dots,1\}$. Each curve has a minimum with cutoff
   radius, the locus of these minima with varying time step
   is tracked by the green dashed line.
   (c)~The relative statistical uncertainty with cutoff radius
   in DMC.
   (d)~The spin-resolved pair correlation functions for same-spin
   ($g_{\uparrow\uparrow}$) and opposite-spin
   ($g_{\uparrow\downarrow}$) electrons for the Coulomb interaction are
   shown in red and those for the pseudopotential are in blue. 
   The dotted blue curve shows the analytic
   correction ($g_{\uparrow\downarrow}^{\text{2body}}$) applied to the cyan
   $g_{\uparrow\downarrow\text{PP}}$ obtained directly from the pseudopotential.
   (e) The error in the energy and speedup obtained with density.  }
 \label{fig:DMC}
\end{figure}

The pseudopotential was calibrated using the exactly soluble two-body
scattering problem and tested against the analytic solution of two electrons
in a parabolic trap.  We now study a system that cannot be solved
analytically: the HEG. We must rely on a numerical approach to determine the
ground state energy, allowing us to expose the computational benefits of
using a pseudopotential. We first study the HEG with DMC as this is the
leading approach for accurate calculations of the ground state
energy~\cite{Ceperley80,Ortiz99,Zong02,Spink_2013}.

We have used the CASINO quantum Monte Carlo code \cite{Needs10} to perform
variational and diffusion Monte Carlo (VMC and DMC) calculations
\cite{Ceperley80,Foulkes_2001}.  The Metropolis algorithm is used in the VMC
method to generate a set of electron configurations distributed according to
the square modulus of the trial wavefunction over which the local energy is
averaged.  In the DMC method, an initial wavefunction is evolved in
imaginary time, which projects out the ground state.  The antisymmetry of
the wavefunction is imposed via the variational fixed-node approximation, in
which the nodal surface remains unchanged during the evolution.  The
simulation proceeds with configurations undergoing drift, diffusion and
birth/death processes, which simulate the evolution of the wavefunction in
imaginary time.  DMC provides an upper bound on the energy that is lower
than the VMC bound calculated with the same trial state.

We focus on a three-dimensional homogeneous electron gas with
$N_{\uparrow}=N_{\downarrow}=57$ electrons and density
$r_{\text{s}}=r_{0}/a_{\text{B}}=2$ with $a_{\text{B}}$ the electron Bohr
radius. The calculation is performed in a periodically repeated simulation
cell and the interaction energy is calculated using Ewald
summation~\cite{Ewald21,Tosi64}. We first construct a variational
wavefunction $\psi=\e{J}D$ that is the product of a Jastrow factor $\e{J}$
and a Slater determinant $D=\hat{\mathcal{A}}\{\prod_{i\in
  N_{\uparrow}}^{\vec{k}\in
  k_{\text{F}}}\e{\cmplxi\vec{k}\cdot\vec{r}_{i}}\}\hat{\mathcal{A}}\{\prod_{i\in
  N_{\downarrow}}^{\vec{k}\in
  k_{\text{F}}}\e{\cmplxi\vec{k}\cdot\vec{r}_{i}}\}$, where
$\hat{\mathcal{A}}$ is the anti-symmetrization operator that accounts for
fermion statistics.  The lowest energy plane-wave states $\vec{k}$ are used
to form the orbitals and periodic boundary conditions are applied.  The log
of the Jastrow factor is
\begin{align}
  J\!=\!\!\!\!\!\!\!\!
  \sum_{\substack{\alpha,\beta\in\{\uparrow,\downarrow\}\\
  i,j\in N_{\alpha},N_{\beta}}}\!\!
  \left[1\!-\!\frac{|\vec{r}_{i}\!-\!\vec{r}_{j}|}{L_{\text{u}}}\right]^{3}
  \!\!\!\Theta(L_{\text{u}}\!-\!|\vec{r}_{i}\!-\!\vec{r}_{j}|)
  \!\sum_{k=0}^{N_{\text{u}}}\!u_{k\alpha\beta}\!\!
  \left[\frac{|\vec{r}_{\!i}\!-\!\vec{r}_{\!j}|}{L_{\text{u}}}\right]^{k}
  \!\!\!\!\punc{,}\nonumber
\end{align}
which includes strongly repulsive electron-electron correlations.  We
describe $J$ by a polynomial expansion of order $N_{\text{u}}=8$ in the
electron-electron separation,~\cite{Drummond04} and $L_{\text{u}}$ is a
cutoff length.  The behavior of the Jastrow factor at electron coalescence
can be fixed by the Kato cusp conditions~\cite{Kato57}; for the Coulomb
potential we can remove the cusp by setting
$u_{1\alpha\beta}=3u_{0\alpha\beta}+1/2$ for antiparallel spins
($\alpha\ne\beta$) and $u_{1\alpha\alpha}=3u_{0\alpha\alpha}+1/4$ for
parallel spins. However, this scheme leaves a remnant discontinuity in the
local energy. On the other hand, the pseudopotential is smooth
at $r=0$, so there we set $u_{1\alpha\beta}=3u_{0\alpha\beta}$ to
ensure that the wavefunction is smooth at electron coalescence. 
The higher order terms in the Jastrow factor $\{u_{i\ge2,\alpha\beta}\}$
provide the freedom to account for longer-ranged correlations.
We also add
backflow correlations in the Slater determinants using the substitution
$\vec{r}_{i}\mapsto \vec{r}_{i}+\sum_{j\in\{N_{\uparrow},N_{\downarrow}\}}
\eta_{ij}(|\vec{r}_{i}-\vec{r}_{j}|)(\vec{r}_{i}-\vec{r}_{j})$ with
$\eta(r)=(1-r/L_{\eta})^{3}\Theta(L_{\eta}-r)\sum_{k=0}^{N_{\eta}}c_{k}r^{k}$,
where $L_{\eta}$ is a cutoff length, and the expansion in variational
parameters $c_{k}$ is up to order $N_{\eta}=8$~\cite{LopezRios06}.  The
variational coefficients $\{u_{k\alpha\beta},c_{k},L_{\text{u}},L_{\eta}\}$
are optimized using VMC
\cite{variance_minimisation_1999,variance_minimisation_2005}.

The VMC wavefunction was used as the trial state for the DMC
calculation. DMC propagates the electrons in time step increments $\tau$
governed by Schr\"odinger's equation in imaginary time. The evolution with a
Coulomb interaction must have a small time step to properly sample the
rapidly changing local energy near the electron cusp~\cite{Needs10}. All DMC
calculations were performed with at least 1000 walkers. We use the
percentage of the correlation energy $E_{\text{C}}$ retrieved as the measure
of the accuracy. There are two main sources of error, firstly the underlying
VMC trial wavefunction is not exact, having a variance in the local energy
$\sigma_{\text{L}}^{2}=\text{var}(\psi^{-1}\hat{H}\psi)$ that introduces a
systematic error in the DMC estimate of the ground state energy of $\Delta
E=a\sigma_{\text{L}}\tau$~\cite{Leszczynski12}, where $a$ is a system
dependent constant. Secondly, because DMC follows a random walk there is a
statistical uncertainty
$\sigma_{\text{E}}=b\sigma_{\text{L}}/(\tau^{1/2}N^{1/2})$, where $b$ is a
system dependent constant, that can be reduced by taking more samples
$N$. Both sources of error increase with the variance of the local
energy~\cite{Lee11}, which for the Coulomb potential is
$\sigma_{\text{L}}=7.6\times10^{-4}E_{\text{C}}$, and for the
pseudopotential (with $c=r_{0}$) is
$\sigma_{\text{L}}=2.4\times10^{-4}E_{\text{C}}$.  Using the pseudopotential
has resulted in a drop in $\sigma_{\text{L}}$ by a factor of $3.2$, which
should reduce both the systematic and statistical errors.  To expose this we
now vary another parameter that enters both sources of error: the time step.

In the upper panel of \figref{fig:DMC}(a) we first examine the systematic
error in the energy.  The extrapolates of the ground state energy to zero
time step for the Coulomb and pseudopotential interactions agree to within
$0.013\%E_{\text{C}}=0.0012\text{\,au}$ per electron~\cite{Lee11}. This is
better than our goal of chemical accuracy of
$1$\,kcal\,mol$^{-1}=0.0016\text{\,au}$ per electron.  Calculation with
the Coulomb interaction and pseudopotential both have the expected $\Delta
E=a\sigma_{\text{L}}\tau$ linear variation of energy with time step, though
the slope for the Coulomb interaction is $3.5$-times as steep as for the
pseudopotential interaction. This is consistent with the Coulomb interaction
having a $\sigma_{\text{L}}$ that is $3.2$-times as large.  Now that we have
confirmed the analytical form for the systematic error in the energy, we
examine the statistical uncertainty that is expected to be
$\sigma_{\text{E}}=b\sigma_{\text{L}}/(\tau^{1/2}N^{1/2})$.  The lower panel
of \figref{fig:DMC}(a) confirms that the statistical error is well-fitted by
a $\tau^{-1/2}$ power law, and that the ratio of the fitting coefficients is
$3.3$, consistent with the expected ratio from the local energy of $3.2$.

With the behavior of both the systematic and statistical errors verified, we
determine the acceleration offered by the pseudopotential. Considering only
the statistical error,
$\sigma_{\text{E}}=b\sigma_{\text{L}}/(\tau^{1/2}N^{1/2})$, to achieve a
target final uncertainty requires a computational effort that scales with
the number of samples as $N\sim\sigma_{\text{L}}^{2}$. The local energy
calculated with the pseudopotential has an error of $\sigma_{\text{L}}$,
which is $3.2$-times smaller than for the Coulomb interaction, resulting in
a 10-times speedup. However, when using the pseudopotential the systematic
error is also reduced, allowing the calculation to be performed at larger
time steps, which will also reduce the statistical error as
$\sim\tau^{-1/2}$.  We consider these effects on an even footing by
combining the systematic and statistical errors in quadrature to give a
total expected error $\Delta E_{\text{tot}}$ in the estimate of the energy
of
\begin{align}
 \Delta E_{\text{tot}}^{2}&=\Delta E^{2}+\sigma_{\text{E}}^2\neweqnline
 &=\sigma_{\text{L}}^2\left(a^2\tau^2+\frac{b^2}{\tau N}\right)\punc{.}
\end{align}
The systematic contribution to the total error grows with time step while
the statistical uncertainty diverges with decreasing time step. The best
compromise between the two can be found by minimizing the error with respect
to time step $\tau$ to yield
\begin{align}
 \min(\Delta E_{\text{tot}})=
 \frac{3^{1/2}b^{2/3}}{2^{1/6}a^{1/3}}\frac{\sigma_{\text{L}}}{N^{1/3}}\punc{.}
\end{align}
If we aim for a particular target total error the computational effort scales with
the number of samples as $N=\sigma_{\text{L}}^{3}$.  The pseudopotential reduces
$\sigma_{\text{L}}$ by a factor of $3.2$, and therefore the pseudopotential
offers a $\sim30$ fold reduction in computational cost while delivering
chemical accuracy.

With the benefits of the pseudopotential established, in \figref{fig:DMC}(b)
we investigate tuning of the pseudopotential cutoff radius. Starting with a
small cutoff radius, the energy has a minimal systematic error at small time
steps, but the calculation with the Coulomb interaction suffers from a large
local energy variance and the error grows rapidly with time step. As the
cutoff is increased the variance in the local energy is reduced and the
finite time step error falls until it is minimal at $c\approx r_{0}$. At
large cutoff radii $c\gtrsim r_{0}$ the interaction potential is
insufficiently accurate to reproduce the correct ground state energy in the
zero-time-step limit. There is now a high probability that three electrons
will be found within the cutoff radius, whereas the pseudopotential was
calibrated for two-body physics. The error therefore increases rapidly,
independently of the time-step adopted. When selecting the cutoff radius one
should also consider the impact of the variance in the local energy on the
statistical uncertainty in the final result. In \figref{fig:DMC}(c) we show
that with increasing cutoff radius the increasingly smooth pseudopotential
leads to a reduction in the relative uncertainty. At $c=r_{0}$ the relative
uncertainty has fallen by the same factor of $\sim3.2$ as shown in
\figref{fig:DMC}(a).

In \figref{fig:DMC}(d) we study the modification of the pair correlation
function arising from the use of the pseudopotential.  The same-spin pair
correlation function from the Coulomb interaction and the pseudopotential
agree within $0.5\%$. The opposite-spin correlation functions are identical
at separations $r\gtrsim c$ where the underlying potentials are
identical. At $r\lesssim c$ the pseudopotential is smaller than the Coulomb
potential, and therefore the corresponding pair correlation function is
larger. However, at small separations two-body physics dominates, and we can
separately calculate the pair correlation function by solving the same
two-body scattering problem that we used to form the original
pseudopotential. This two-body solution can be used to correct the many-body
estimate of the pair correlation function for the incorrect two-body
effects, bringing it into agreement with the solution for the Coulomb
potential to within $1\%$. Any further deviation can be ascribed to three-
and higher-body physics that occurs for $r\le c$, which is rare as the
electrons are simultaneously Pauli blocked and repelled by the strong
Coulomb repulsion.

Having confirmed the utility, robustness, and accuracy of the
pseudopotential for the electron gas with $r_{\text{s}}=2$ we study the
accuracy of the pseudopotential for electron gases with densities in the
range $1\le r_{\text{s}}\le16$. With the cutoff radius at each density set
according to $c=r_{0}=a_{\text{B}}r_{\text{s}}$, we compare the ground state
energy from the pseudopotential with that of the Coulomb interaction.  In
\figref{fig:DMC}(e) we see that the pseudopotential is able to deliver
ground state energies to better than chemical accuracy with a speedup by a
factor of $\sim30$ across a broad range of densities.

\section{HEG with configuration interaction}

\begin{figure}
 \begin{tabular}{ll}
  (a) Coulomb wavefunction&(b) Pseudopot. wavefunction\\
  \includegraphics[width=0.48\linewidth]{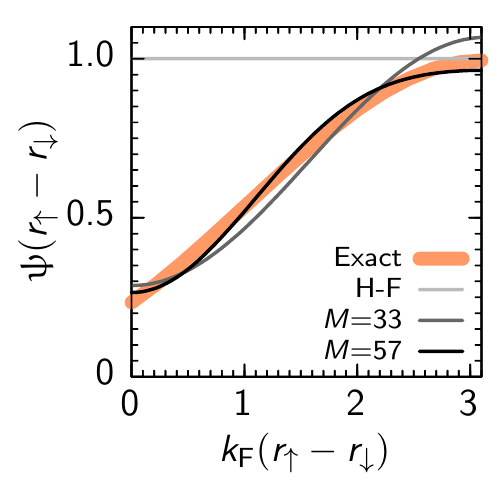}&
  \includegraphics[width=0.48\linewidth]{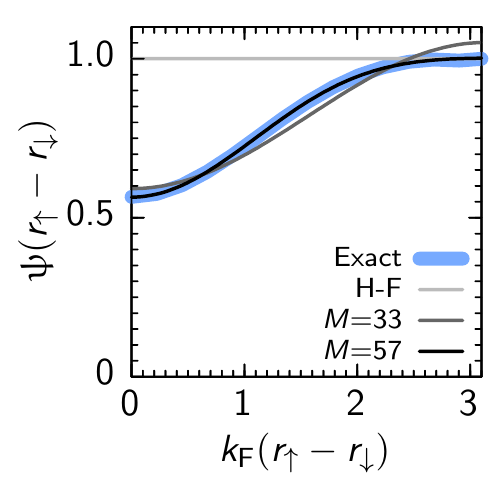}\\
  (c) Wavefunction convergence&(d) Energy convergence\\
  \includegraphics[width=0.48\linewidth]{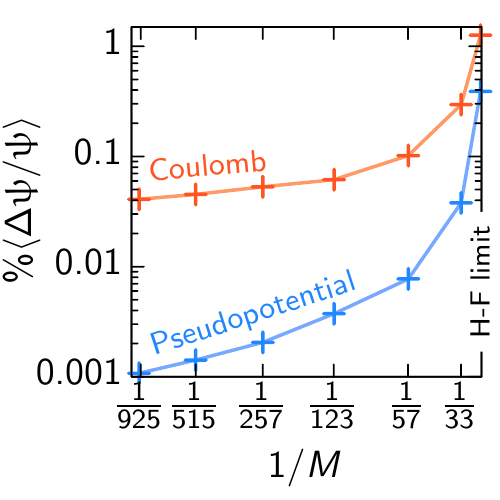}&
  \includegraphics[width=0.48\linewidth]{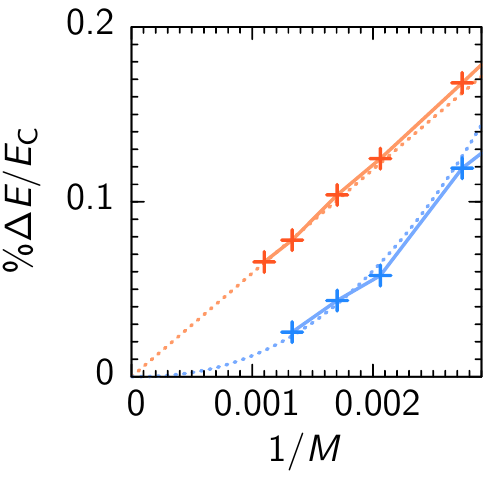}
 \end{tabular}
 \caption{(Color online)
   (a,b) The CID wavefunction for opposite-spin electrons
   passing through coalescence for the HEG at $r_{\text{s}}=2$ with
   increasing plane-wave orbital basis sets of size $M$. The exact solution
   is shown in red (blue) for the Coulomb interaction potential and the
   pseudopotential.
   (c) The average relative error in the wavefunction, and
   (d) the percentage error in the energy with basis set size for the
   Coulomb potential (red) and pseudopotential (blue) with dotted trend
   lines.
 }
 \label{fig:CI}
\end{figure}

The success of the pseudopotential for studying the HEG with DMC motivates
us to consider a second complementary approach to examine the HEG,
Configuration Interaction Doubles (CID)~\cite{Hylleraas28,Meyer76}. We adopt
a plane-wave basis for our CID calculations, which offers a robust test of
the portability of the pseudopotential. CID theory starts from the
Hartree-Fock ground state and includes electron correlations through double
excitations into the unoccupied (plane-wave) orbitals. In the Coulomb
potential, the wavefunction has a gradient discontinuity at
electron-electron coalescence that must be described by a large number $M$
of plane-wave basis states with a computational cost that scales as
$\mathcal{O}(M^6)$. However, the pseudopotential removes the
electron-electron cusp rendering the wavefunction smooth, which therefore
should require fewer plane waves to describe the ground state and in turn
reduce the computational expense.

The major computational gain offered by the pseudopotential is to aid the
description of the behavior at electron coalescence, and therefore we first
examine how the wavefunction at coalescence of two opposite-spin electrons
evolves with the size of the plane-wave basis set. In the presence of the
Coulomb interaction we compare the exact relative wavefunction with that
from a finite basis set in \figref{fig:CI}(a). The Hartree-Fock wavefunction
does not include opposite-spin correlations, and therefore the relative
wavefunction is constant at electron coalescence.  The description of the
gradient discontinuity in the wavefunction at coalescence improves with
increasing basis set size.  In \figref{fig:CI}(b) we repeat the exercise in
the presence of the pseudopotential. The pseudopotential has zero gradient
at $r=0$ so the exact wavefunction is now smooth at electron
coalescence. This allows the shape of the wavefunction to be described
accurately by a relatively small basis set. To quantify the change in
wavefunction with basis set size, we examine in \figref{fig:CI}(c) the
relative error in the wavefunction, spatially averaged within the exchange
correlation hole, $k_{\text{F}}r\le\pi$, using
\begin{align}
 \left\langle\frac{\Delta\psi}{\psi}\right\rangle^2=
 \frac{3k_{\text{F}}^3}{\pi^3}\int_{0}^{\pi/k_{\text{F}}}
 \left(1-\frac{\psi_{M}(r)}{\psi_{\infty}(r)}\right)^2
 4\pi r^2\diffd r\punc{,}
\end{align}
where $\psi_{M}(r)$ is the relative wavefunction on coalescence of two
opposite-spin electrons with separation $r$, calculated with CID and a basis
set size $M$.  The average error for the Coulomb potential falls slowly with
increasing basis set size. However, a proper description of the wavefunction
at coalescence requires a plane-wave basis set with a wave vector of at
least $\sim1.5k_{\text{F}}$, corresponding to a basis set size of
$M=57$. \figref{fig:CI}(c) shows that here the error in the wavefunction
drops markedly and the wavefunction is over ten times more accurate than
that for the Coulomb interaction at the same basis set size. With large
basis sets the wavefunction obtained with the pseudopotential converges more
rapidly than that for the Coulomb interaction.

Now that we have shown that the pseudopotential facilitates CID calculations
of the wavefunction we study the impact on evaluating the ground state
energy. Both estimates tend towards the same ground state energy, confirming
the accuracy of the pseudopotential.  In \figref{fig:CI}(d) we show that the
error in the ground state energy calculated with the Coulomb interaction
scales as $1/M$~\cite{Gruneis12} whereas with the pseudopotential it scales
as $1/M^{7/3}$, which is the same improvement as seen with explicitely
correlated methods~\cite{Gruneis12}.  The pseudopotential delivers benchmark
chemical accuracy of $0.017\%E_{\text{C}}=0.0016\text{\,au}$ per electron
with a $\sim50\%$ smaller basis set and, since the computational cost of CID
scales as $\mathcal{O}(M^6)$, this corresponds to a speed-up of a factor of
$\sim32$. Even greater computational gains could be expected at higher
levels of target accuracy.

The pseudopotential has contributed to reducing the basis set size required
in a CID calculation. This benefit is expected to be carried over to
more accurate configuration interaction approaches, for example coupled
cluster that overcomes the errors introduced into CID by unlinked
diagrams~\cite{Bartlett07}. Here we adopted a plane-wave basis set,
however applications of configuration
interaction to molecules often express the
wavefunction in a coordinate basis set centered
on the atoms. The pseudopotential takes a smooth polynomial form so the
two-electron integrals could be evaluated efficiently as summations over the
Boys function~\cite{Reine12}.

\section{Lithium \& Beryllium atoms}

\begin{figure}
 \begin{tabular}{ll}
  (a) Atom energy&(b) Ion energy\\
  \includegraphics[clip,width=0.48\linewidth]{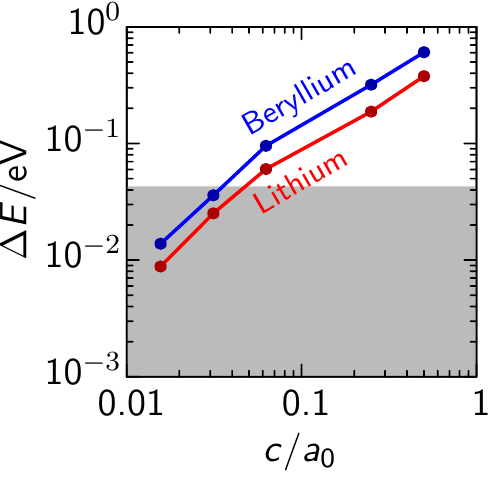}&
  \includegraphics[clip,width=0.48\linewidth]{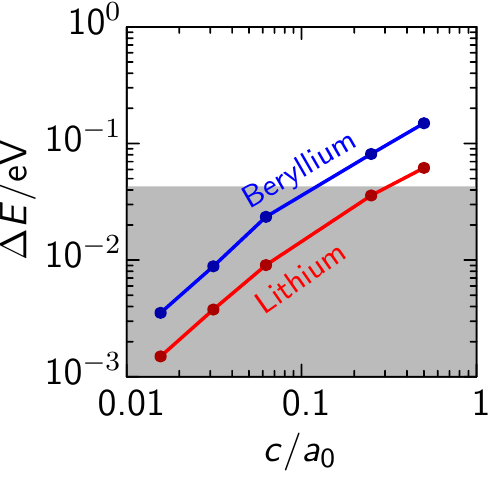}\\
  (c) Ionization energy&(d) Speedup\\
  \includegraphics[clip,width=0.48\linewidth]{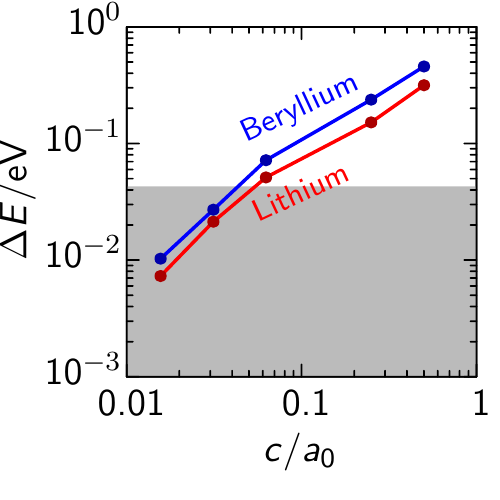}&
  \includegraphics[clip,width=0.48\linewidth]{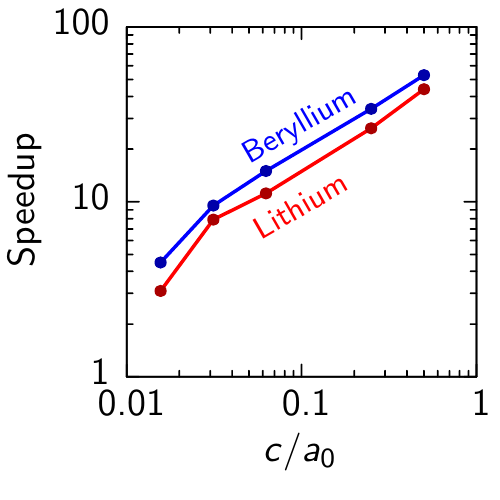}
 \end{tabular}
 \caption{(Color online)
  (a) Error $\Delta E$ in the total energy with cutoff radius. The red points
  are for the isolated Li atom and blue are for the Be atom. The gray shading
  denotes where the results attain chemical accuracy. 
  (b) The error in the total energy of the Li (red) and
  Be (blue) ions.
  (c) The error in the ionization energy of a Li (red) and 
  Be (blue) atom.
  (d) The speedup of the DMC calculation. The red points
  show the Li atom and blue the Be atom.
}
 \label{fig:lithium}
\end{figure}

The HEG is arguably the most important model of interacting
electrons. However, in real systems, the background charge density due to
the atomic nuclei is non-uniform and so the electron density varies in
space.  In order to study the performance of the pseudopotential in an
inhomogeneous system, we perform DMC calculations of the energy of the
lithium and beryllium atoms. These atoms are simple real-life systems that
could expose errors introduced by three-body scattering. Accurate reference
results from analytic integration and recursion
relations~\cite{Puchalski08,Chakravorty93} are also available, making these
systems an ideal test bed for evaluating the performance of the
electron-electron pseudopotential.


The trial wavefunction is constructed from single-particle orbitals 
in a Gaussian basis set generated by an all-electron calculation
performed using CRYSTAL~\cite{Dovesi06}.
The trial
wavefunction consists of a determinant of DFT orbitals multiplied by a
Jastrow correlation factor. The parameters in the Jastrow factor are
optimized using a variance minimization technique~\cite{Drummond04}.  The
optimized VMC wavefunction is used as a starting point for a DMC
calculation.

We first study a solitary Li atom, containing one down-spin and two up-spin
electrons. We present our estimates for the ground state energy in
electronvolts for ready comparison with the real-life system. In
\figref{fig:lithium}(a) we show the variation of the accuracy compared with
the pure Coulomb interaction. The energy for the exact Coulomb system,
$-203.379$\,eV, agrees with reference results from analytic integration and
recursion relations~\cite{Puchalski08,Chakravorty93} within $0.024$\,eV per
atom.  The error decreases as the cutoff radius is reduced. If we aim for an
error of order chemical accuracy ($0.025$\,eV per atom) we require
$c\lesssim0.03a_{0}$. \figref{fig:lithium}(d) shows that relative to the
calculation with the Coulomb interaction, the smoother pseudopotential
reduces the local variance and therefore accelerates the calculation, with
greater effect for larger cutoff radii. The pseudopotential offers a speedup
by a factor of $\sim5$ while still attaining chemical accuracy.

The results for the Be atom follow the same trend as for the Li atom. For
the Be atom we predict a ground state energy of $-398.932$\,eV per atom,
again within $0.020$\,eV of reference results from analytic integration and
recursion relations~\cite{Puchalski08,Chakravorty93}.  The pseudopotential
performs slightly better for the Be than the Li atom, possibly due to the
increased prevalence of electron-electron relative to electron-ion
interaction terms. We also determine the energy of the Li$^{+}$ and Be$^{+}$
ions in \figref{fig:lithium}(b).  The error is now significantly reduced due
to the removal of the three-body error for Li$^{+}$, and its reduction for
Be$^{+}$.  The growth of the error in the energy estimate is similar to that
for the Li and Be atoms.  This means that in \figref{fig:lithium}(c) the
magnitude of the error in the ionization energy grows with cutoff radius. We attain
chemical accuracy ($0.025$\,eV per atom) at $c\lesssim0.05a_{0}$.

For a fixed target accuracy the speedup of the pseudopotential calculation
for the Li and Be atoms is smaller than for the HEG. This is because in the
HEG we focused on the error per electron, whereas here we focus on the error
per atom, which includes three or four electrons, therefore inflating the
error. However, even if we ignore this, the electron-electron
pseudopotential offers a $5$-times acceleration for high accuracy work,
whereas for example, for high throughput structure prediction calculations
an order of magnitude less accuracy is required~\cite{Pickard11} so a
pseudopotential would offer a $50$-times speedup. For a molecule chemical
accuracy typically relates to the energy difference between two configurations
rather than total energy for which the pseudopotential is expected to be more
accurate.

\section{Discussion}

We have developed a pseudopotential for the repulsive Coulomb
interaction. The pseudopotential delivers accurate scattering states for
incident wave vectors and angular momentum channels found in an electron
gas, while its smoothness accelerates computation. With the cutoff radius
set to the typical electron separation the pseudopotential delivers the
correct many-body physics, and within the cutoff radius two-body physics
dominates where predictions for the exchange correlation hole can be
corrected analytically. The cutoff radius can be reduced to zero, making the
pseudopotential systematically improvable. The pseudopotential was shown to
deliver chemical accuracy for the HEG and to accelerate both the DMC and CID
methods by a factor of $\sim30$. The pseudopotentials were also shown to
accelerate the calculation of the isolated lithium and beryllium atom by a
factor of $5$ for high accuracy work, and in situations where lower accuracy
is required, for example high throughput structure prediction calculations,
the pseudopotentials would provide a $50$-times acceleration.

The performance and simplicity of the electron-electron pseudopotential
makes it portable across many-body techniques such as VMC, DMC, truncated
CI, coupled cluster theory, and M{\o}ller-Plesset theory. The formalism
developed can be applied more widely in scattering problems in condensed
matter to develop pseudopotentials for dipolar interactions and also the
contact interactions found in atomic gases~\cite{Bugnion14}. The approach
can also be applied to classical physics, for example the Coulomb
interaction studied here has the same force law as Newtonian gravity used in
simulations of galactic dynamics~\cite{Binney08}. Here a pseudopotential
could overcome the high computational cost and correctly capture the motion
of stars during close encounters.

\acknowledgments {The authors thank George Booth and Pascal Bugnion for
  useful discussions, and GJC acknowledges the financial support of the
  Royal Society and Gonville \& Caius College.}

\end{document}